# STATIC MAGNETIC DIPOLE SOLUTIONS OF EINSTEIN – MAXWELL EQUATIONS FROM THE SOLITON SOLUTIONS OF A KERR OBJECT

**A. CHAUDHURI***

Department of Physics, B. K. C. College, 111/2, B.T. Road, Bon Hooghly, Kolkata-700108, INDIA

Email: (i) avinandac@gmail.com , (ii) aviphys@gmail.com ,

**S. Nandi**

Rajkhamar High School, Rajkhamar, Dist.- Bankura (W.B.), - 722205, INDIA

Email : sprynandi@gmail.com

and

**S. CHAUDHURI**

Chaudhuri Lane, R.K. Pally, Kalna Road, Burdwan – 713101, INDIA

Email: schaudhuri.phys@gmail.com

( * Corresponding author )

## ABSTRACT

Using the solution generating techniques of Das and Chaudhuri (Pramana J Phys **40**:277, 1993; Pramana J Phys **58**:449, 2002), static magnetovac solutions of Einstein-Maxwell equations in general relativity are constructed from the stationary gravitational soliton solutions of Einstein's field equations corresponding to a Kerr object. The techniques followed in the present paper have not been much used in the literature so far although the results are obtained in a straightforward way. The stationary gravitational two-soliton solutions of Einstein's field equations for a Kerr object are generated using the soliton technique of Belinskii and Zakharov (Sov. Phys. JETP **48**:985, 1978; Sov. Phys. JETP **50**:1, 1979). In the next step, following the procedure of Das and Chaudhuri, mentioned above, static magnetovac solutions of Einstein – Maxwell field equations corresponding to the generated stationary two-soliton solutions of the Kerr object are constructed. The solutions are found to be well behaved at spatial infinity and contain monopole, dipole and other higher mass multipoles. The mass and the dipole moment of the source are evaluated. It is shown that by redefining some constants appearing in the solutions, Bonnor's magnetic dipole solutions (Z.Phys. **190**: 444, 1966) are faithfully reproduced.

# 1   INTRODUCTION

In the paper, we construct static magnetic dipole solutions of Einstein – Maxwell (EM) field equations from the stationary gravitational two-soliton solutions of a Kerr object [1]. In order to generate magnetostatic solutions from stationary gravitational two-soliton solutions, we used two transformation techniques, already known in the literature, developed by Das and Chaudhuri [2, 3]. We followed two steps: (i) stationary gravitational two-soliton solutions corresponding to a Kerr object are constructed using the soliton technique of Belinskii and Zakharov [BZ] [4, 5] and (ii) the generated stationary gravitational solutions in (i) are then used to construct magnetostatic solutions of EM field equations following the simple transformations presented by Das and Chaudhuri [2, 3]. The mass and the magnetic dipole moment of the source are evaluated and the solutions are found to be asymptotically flat containing monopole, dipole and other higher mass multipoles. For a particular value of the magnetic parameter the mass of the source becomes zero and one obtains the dipole moment of a mass less source. This represents finite current loop at large distances. Finally, by suitable readjustment of the constants, it is shown that the generated magnetovac solutions exactly coincide with the magnetic dipole solutions obtained by Bonnor [6].

Chaudhuri et.al [7] first used the techniques discussed in Refs. [2, 3] to generate magnetostatic solutions from stationary gravitational two-soliton solutions. They considered a Laplace's solution as seed in the general static metric and derived the stationary gravitational field. The corresponding magnetostatic solutions are then constructed using the transformations [2, 3]. In the present paper we have used the Kerr metric, obtained by the soliton technique of Belinskii and Zakharov [4, 5], as seed and static magnetic dipole solutions are presented. Chaudhuri et.al [8] also constructed static magnetovac solutions of EM field equations from the stationary gravitational fields using the technique discussed in Ref. [2]. In their works they used the method of Manko et.al [9, 10, 11] to generate the stationary gravitational fields. The results obtained in [8] and that in the present paper have some resemblance although the stationary gravitational fields are derived using different methods. It may be pointed out that in Ref. [8] one does not have to use the transformation presented in Ref. [3] where as in the present paper the transformation [3] has to be applied to obtain the static magnetic solutions. The reason for which is discussed below.

In our work the generation of the magnetostatic solutions is based on the comparison of the metric coefficients of stationary and static fields. The essence of the technique is: we have constructed the stationary gravitational field corresponding to the Kerr object using the soliton technique of BZ [4, 5]. Since it is difficult to relate directly the metric coefficients arising from soliton technique to those obtained from magnetostatic solutions of EM field equations, we have to use a second transformation discussed in Ref. [3] to generate magnetostatic potential and hence the dipole moment of the source. It is worth-while to mention that although the transformations [2, 3] are known earlier, the method followed in the paper, particularly the use of the second transformation [3], to generate magnetostatic solutions from stationary gravitational two-soliton solutions are not much found in the literature although it makes the solution generation process much simpler. The results are obtained in a straightforward way. Probably the application of these techniques mentioned above escaped our attention so far.

The following sections are divided as: in Sect. 2, the method of BZ [4, 5] is described in brief. Sect. 3 contains the generation of stationary gravitational solutions of the Kerr object [1] using the soliton technique of BZ. The transformation techniques (Refs. [2, 3]) and the constructed magnetostatic solutions are presented in sect. 4. Our conclusion follows in Sect. 5.

## 2 SOLUTION GENERATING TECHNIQUE OF BELINSKII AND ZAKHAROV:

The 'inverse scattering method' (ISM), popularly known as the *'soliton technique',* discovered by Blenskii and Zakharov (BZ) [4, 5] has wide applications of generating exact solutions of Einstein's field equations in general relativity. In this section we describe the method of BZ in brief mentioning only the basic equations required for our purpose (because of the fact that the soliton technique of BZ is well known and well discussed in the literature of general relativity). A detailed discussion on the soliton technique, to name only a few, can be found in Refs. [ 4, 5, 12 – 17].

Consider a stationary axially symmetric metric in cylindrical coordinates *(r, z)*,

$$ds^2 = g_{ab}dx^a dx^b + e^\nu (dr^2 + dz^2), \qquad (1)$$

where the indices $a, b$ take values 1, 2 and $t, \phi = x^1, x^2$ respectively. The metric coefficients $g_{ab}$ and $v$ are functions of $r, z$ only.

Physically acceptable solutions corresponding to metric (1) have to satisfy, without loss of generality, the additional condition [13]

$$\det g_o = -r^2, \tag{2}$$

where $g_0$ is a (2 x 2) matrix associated with $g_{ab}$.

Now define the following two quantities $U$ and $V$ as

$$U = r g_r g^{-1} \quad \text{and} \quad V = r g_z g^{-1}, \tag{3}$$

where the subscripts $r, z$ represent partial differentiations.

With the supplementary condition given in Eq. (2), the Einstein's field equations for the metric (1) can be written in the form [5]

$$U_r + V_z = 0, \tag{4}$$

$$Tr(U^2 - V^2) = 4(1 + r v_r), \tag{5}$$

$$Tr(UV) = 2 r v_z. \tag{6}$$

It is evident that for a known $g$, the metric coefficient $v$ can be derived from eqs. (5) and (6). Equation (4) is the integrability condition for $v$.

According to BZ, the soliton solutions to equation (4) are obtained from the solutions of the following pair of differential equations [4, 5, 13]

$$\left( \partial_r + \frac{2 \lambda r}{\lambda^2 + r^2} \partial_\lambda \right) \Omega = \frac{rU + \lambda V}{\lambda^2 + r^2} \Omega, \tag{7}$$

$$\left( \partial_z - \frac{2 \lambda^2}{\lambda^2 + r^2} \partial_\lambda \right) \Omega = \frac{rV - \lambda U}{\lambda^2 + r^2} \Omega, \tag{8}$$

where $\lambda$ is a complex spectral parameter and $\Omega$, a 2x2 matrix, is called the eigenfunction of the equations (7) and (8). The eigenfunction $\Omega$ is a function of (r, z, $\lambda$) i.e. $\Omega = \Omega, (r, z, \lambda)$. Furthermore, $\Omega, (r, z, \lambda)$ satisfies the relation [13]

$$\Omega(r, z, \lambda)\big|_{\lambda=0} = g_0(r, z), \tag{9}$$

The new metric coefficients are constructed from the following relations [5,13,16]

$$g'_{ab} = (g_o)_{ab} - \sum_{k,l=1}^{N} N_a^l (\Gamma^{-1})_{lk} N_b^k \mu_k^{-1} \mu_l^{-1}, \tag{10}$$

$$\Gamma_{kl} = m_c^k (g_o)_{ca} m_a^l (\mu_k \mu_l + r^2)^{-1}, \tag{11}$$

$$m_a^k = m_{oc}^k \left[\Omega^{-1}(r, z, \lambda = \mu_k)\right]_{ca}, \tag{12}$$

$$N_a^k = m_b^k (g_o)_{ba}, \tag{13}$$

and the pole trajectories $\mu_k$'s are defined by

$$\mu_k = \omega_k - z + \left[(\omega_k - z)^2 + r^2\right]^{1/2}, \tag{14}$$

where $m_{oc}^k$ and $\omega_k$ are arbitrary constants. The indices $k$ and $l$ can have values from 1 to $N$. Here $N$ is the number of solitons i.e. the number of poles that appears in the scattering matrix. Solutions are characterized by the value of $N$ such that for $N = 1$, it is called one-soliton solutions, $N = 2$, is the two-soliton solutions etc.

The constructed metric $g'_{ab}$ in equation (10) satisfies equation (4). However, in general, it does not satisfy equation (2). One thus has to search for the physically acceptable solution $g = g^{ph}_{ab}$; which is defined by the relation [ 13 ]

$$g^{ph}_{ab} = -r(-\det g')^{-1/2} g'_{ab}, \tag{15}$$

where, $\det g' = (-1)^N r^{2N} \left(\prod_{k=1}^{N} (\mu_k^2)^{-1}\right) \det g_o. \tag{16}$

Letelier has shown that for general N-soliton solutions the function $v$ is given by [13]

$$v_N = v_0 + \ln\left[ r^{-N^2/2} \left(\prod_{k=1}^{N}\mu_k\right)^{N+1} \left(\prod_{\substack{k,l=1\\k>l}}^{N}(\mu_k - \mu_l)^2\right)^{-2} \det \Gamma_{kl}\right] + \ln C_N, \qquad (17)$$

where $v_0$ is the $v$–function of the static metric and $C_N$ are arbitrary constants. The expression for $v_0$ has been given in the next section in equation (20).

### 3 KERR SOLUTION USING SOLITON TECHNIQUE

A general static axially symmetric metric in $(r, z)$ coordinates is represented by

$$ds^2 = e^{v_0}(dr^2 + dz^2) + r^2 e^{-\psi} d\phi^2 - e^{\psi} dt^2, \qquad (18)$$

where one of the metric coefficients $\psi = \psi(r, z)$ satisfies Laplace's equation

$$\psi_{rr} + \frac{1}{r}\psi_r + \psi_{zz} = 0, \qquad (19)$$

and the other metric coefficient $v_0$ is obtained from the Laplace's solution $\psi$ by the relation [13]

$$v_0(\psi) = -\psi + \frac{1}{2}\int\left[ r\left(\psi_r^2 - \psi_z^2\right)dr + 2\psi_r \psi_z dz\right], \qquad (20)$$

The relations (4) – (6) reduces to equations (19) and (20) for the metric (18).

We take the 'seed' $g_0$ corresponding to metric (18) as

$$g_o = \begin{pmatrix} -e^{\psi} & 0 \\ 0 & r^2 e^{-\psi} \end{pmatrix}, \qquad (21)$$

which satisfy equation (2) i.e. $\det g_o = -r^2$.

Accordingly the eigenfunction $\Omega$ which is also a 2x2 diagonal matrix has to satisfy equation (9) and we assume $\Omega$ as

$$\Omega = \begin{pmatrix} -e^{F_k} & 0 \\ 0 & (r^2 - 2\lambda z - \lambda^2)e^{-F_k} \end{pmatrix}, \tag{22}$$

where the function $F_k = F_k(r, z, \lambda)$ and $F_k = \psi$ when $\lambda = 0$.

The relations satisfied by $F_k$ are [13]

$$\left( r\frac{\partial}{\partial r} - \lambda\frac{\partial}{\partial z} + 2\lambda\frac{\partial}{\partial \lambda} \right) F_k = r\frac{\partial \psi}{\partial r},$$

$$\left( r\frac{\partial}{\partial \lambda} + \lambda\frac{\partial}{\partial r} \right) F_k = r\frac{\partial \psi}{\partial z}, \tag{23}$$

We now use prolate spheroidal coordinates $(x, y)$ and take the Laplace's solution $\psi$ as

$$\psi = \frac{\alpha_0}{x + y}, \qquad (\alpha_0 \text{ is a constant}), \tag{24}$$

to generate stationary gravitational solutions using the method of BZ.

Prolate spheroidal coordinates $(x, y)$ are related to $(r, z)$ coordinates by [16]

$$r^2 = K^2(x^2 - 1)(1 - y^2), \qquad z = z_1 + Kxy, \tag{25}$$

where, $K$ and $z_1$ are constants.

In soliton technique the solutions are required along the pole trajectories $\lambda = \mu_k$, hence for two-soliton solutions there exist two poles $\mu_1$ and $\mu_2$ and accordingly $F_k$, should have two values $F_1$ and $F_2$.

We take the arbitrary constant $\omega_k$ in equation (14) as real with values $\omega_1$ and $\omega_2$ for two-soliton solutions and obtain from eqs. (14) and (25) the values of $\mu_1$ and $\mu_2$ as

$$\mu_1 = K(x+1)(1-y),$$

$$\mu_2 = K(x-1)(1-y), \tag{26}$$

It has been shown that $F_1$ and $F_2$ given by equation (23) satisfy the following relations in prolate spheroidal coordinates $(x, y)$ as [16],

$$2(x － y)F_{1,x} ＝ (1 ＋ y)[(x － 1)\psi_{,x} ＋ (1 － y)\psi_{,y}],$$

$$2(x ＋ y)F_{2,x} ＝ (1 ＋ y)[(x ＋ 1)\psi_{,x} ＋ (1 － y)\psi_{,y}],$$

$$2(x － y)F_{1,y} ＝ (x － 1)[(1 ＋ y)\psi_{,y} － (x ＋ 1)\psi_{,x}],$$

$$2(x ＋ y)F_{2,y} ＝ (x ＋ 1)[(1 ＋ y)\psi_{,y} － (x － 1)\psi_{,x}]. \tag{27}$$

Where a 'comma' represents partial differentiation with respect to the indices as indicated.

With $\psi$ given by equation (24), it is a simple task to find $F_1$ and $F_2$ from eq. (27) and we have

$$F_1 ＝ \frac{1}{2}\alpha_0 (1 ＋ y)(x ＋ y)^{-1} \tag{28}$$

$$F_2 ＝ \frac{1}{2}\alpha_0 (1 ＋ y)(x ＋ 1)(x ＋ y)^{-2} \tag{29}$$

The metric coefficients $g_{ab}{}^{ph}$ are then determined from equations (10) – (15), (21) – (22), (26), (28) – (29). With the substitutions

$$c_1 ＝ m^1{}_{02} m^2{}_{02} (4\omega_1 \omega_2)^{-1}, \qquad c_2 ＝ m^1{}_{01} m^2{}_{01},$$

$$c_3 ＝ -m^1{}_{02} m^2{}_{01} (2\omega_1)^{-1}, \qquad c_4 ＝ -m^2{}_{02} m^1{}_{01} (2\omega_2)^{-1}, \tag{30}$$

physically acceptable metric coefficients are found to be

$$g_{11}{}^{ph} ＝ -\exp\left[\frac{\alpha_0}{x ＋ y}\right] \begin{bmatrix} c_1{}^2 (1 － y^2) e^{-2k_1} ＋ c_2{}^2 (1 － y^2) e^{2k_1} － c_3{}^2 (x^2 － 1) e^{-2k_2} \\ － c_4{}^2 (x^2 － 1) e^{2k_2} ＋ 2 c_1 c_2 (x^2 － y^2) \end{bmatrix} L^{-1}$$

$$\tag{31}$$

$$g_{22}{}^{ph} ＝ K^2 (x^2 － 1)(1 － y^2) \exp\left[-\frac{\alpha_0}{x ＋ y}\right].$$

$$\begin{bmatrix} c_1{}^2 (1 ＋ y)^3 (1 － y)^{-1} e^{2k_1} ＋ c_2{}^2 (1 － y)^3 (1 ＋ y)^{-1} e^{-2k_1} － c_3{}^2 (x － 1)^3 (x ＋ 1)^{-1} e^{2k_2} \\ － c_4{}^2 (x ＋ 1)^3 (x － 1)^{-1} e^{-2k_2} ＋ 2 c_1 c_2 (x^2 － y^2) \end{bmatrix} L^{-1}$$

$$g_{12}{}^{ph} = g_{21}{}^{ph} = 2KL^{-1} \begin{bmatrix} (x+y)\{c_1 c_3 (x-1)(1+y)e^{k_3} - c_2 c_4 (x+1)(1-y)e^{-k_3}\} \\ -(x-y)\{c_1 c_4 (x+1)(1+y)e^{k_4} - c_2 c_3 (x-1)(1-y)e^{-k_4}\} \end{bmatrix} \quad (33)$$

$$v_2 = v_0 + \ln L - \ln\left[16K^2(x^2 - y^2)\right] + \ln C_2 , \quad (34)$$

$$v_0 = -\frac{\alpha_0}{(x+y)}\left[1 + \frac{1}{2}\alpha_0(x^2 - 1)(1 - y^2)(x+y)^{-3}\right] \quad (35)$$

$$L = c_1^2(1+y)^2 e^{2k_1} + c_2^2(1-y)^2 e^{-2k_1} + c_3^2(x-1)^2 e^{-2k_2} + c_4^2(x+1)^2 e^{2k_2} - 2c_1 c_2(x^2 - y^2) ,$$

(36)

$$k_1 = \alpha_0 \frac{(1 + 2xy + y^2)}{2(x+y)^2} , \quad (37)$$

$$k_2 = \alpha_0 \frac{(1-y^2)}{2(x+y)^2} , \quad (38)$$

$$k_3 = (k_1 - k_2) ,$$

$$k_4 = (k_1 + k_2) , \quad (39)$$

The stationary metric (1) is thus constructed in equations (31) - (39).

Kerr metric is obtained from the generated solutions, if one substitutes, $\alpha_0 = 0$ and $c_1 = c_2$ in eqs. (31) – (39). The metric coefficients are

$$g_{11}^{ph}(Kerr) = -\frac{[2c_1^2\{x^2 + (1 - 2y^2)\} - (c_3^2 + c_4^2)(x^2 - 1)]}{[2c_1^2\{x^2 - (1 + 2y^2)\} - c_3^2(x-1)^2 - c_4^2(x+1)^2]}$$

(40)

$$g_{22}^{ph}(Kerr)$$

$$= -K^2 \frac{[2c_1^2(x^2-1)\{(x^2+3y^2)+(1+y^2)^2\}-(1-y^2)\{c_3^2(x-1)^4+c_4^2(x+1)^4\}]}{[2c_1^2\{x^2-(1-2y^2)\}-c_3^2(x-1)^2-c_4^2(x+1)^2]}$$

$$g_{12}^{ph}(Kerr) = g_{21}^{ph}(Kerr) = 4K \frac{[c_1 c_3 \{x^2 - x(1-y^2) - y^2\} - c_1 c_4 \{x^2 + x(1-y^2) - y^2\}]}{[2c_1^2 \{x^2 - (1-2y^2)\} - c_3^2 (x-1)^2 - c_4^2 (x+1)^2]} \qquad (41)$$

(42)

With further substitution of $c_1 = h$, $c_3 = h^2$, $c_4 = 1$, ($h$ being a constant), in equation (40) the Kerr metric takes its familiar form

$$g_{11}{}^{ph}(Kerr) = -\frac{(p^2 x^2 + q^2 y^2 - 1)}{(px+1)^2 + q^2 y^2}, \qquad (43)$$

where $p$ and $q$ are defined by

$$p = \frac{(1-h^2)}{(1+h^2)} \quad \text{and} \quad q = \frac{2h}{(1+h^2)}$$

such that $\quad p^2 + q^2 = 1$, \hfill (44)

.

## 4  GENERATION OF MAGNETOSTATIC SOLUTIONS:

In this section, magnetostatic solutions are generated from the two-soliton solutions corresponding to the Kerr metric. We first describe in brief the transformations presented in Refs. [2, 3] and then construct the magnetovac solutions.

For an axially symmetric stationary space-time in cylindrical coordinates ($r$, $z$) given by [18]

$$ds^2 = e^u (dt - \omega\, d\phi)^2 - e^{-u} \left[ e^{2\gamma^s} (dr^2 + dz^2) + r^2 d\phi^2 \right], \qquad (45)$$

the vacuum Einstein's field equations are [18, 2]

$$u,_{rr} + u,_{zz} + r^{-1} u,_r = -e^{2u} r^{-2} (\omega,_r^2 + \omega,_z^2), \qquad (46)$$

$$\omega,_{rr} + \omega,_{zz} - r^{-1} \omega,_r = -2(u,_r \omega,_r + u,_z \omega,_z), \qquad (47)$$

$$4\gamma^s_{,r} = r(u^2_{,r} - u^2_{,z}) - e^{2u}r^{-2}(\omega^2_{,r} - \omega^2_{,z}) \tag{48}$$

$$2\gamma^s_{,z} = r[u_{,r}\, u_{,z} - e^{2u}r^{-2}\omega_{,r}\, \omega_{,z}], \tag{49}$$

where the metric coefficients $u$, $\omega$ and $\gamma^s$ are functions of $r$ and $z$ only. The superscript $s$ refers to stationary case and a comma represents partial differentiation.

On the other hand, an axially symmetric static metric in $(r, z)$ coordinates is given by [18]

$$ds^2 = e^{2\beta}dt^2 - e^{-2\beta}\left[e^{2\gamma^m}(dr^2 + dz^2) + r^2 d\phi^2\right]. \tag{50}$$

The metric coefficients $\beta$ and $\gamma^m$ are functions of $r, z$ only.

The EM field equations corresponding to the line element (50) are [18, 19]

$$\beta_{,rr} + r^{-1}\beta_{,r} + \beta_{,zz} = e^{2\beta}r^{-2}(A^2_{3,r} + A^2_{3,z}) \tag{51}$$

$$A_{3,rr} - r^{-1}A_{3,r} + A_{3,zz} = -2(\beta_{,r}\, A_{3,r} + \beta_{,z}\, A_{3,z}) \tag{52}$$

$$\gamma^m_{,r} = r(\beta^2_{,r} - \beta^2_{,z}) + e^{2\beta}r^{-2}(A^2_{3,r} - A^2_{3,z}) \tag{53}$$

$$\gamma^m_{,z} = 2r[\beta_{,r}\, \beta_{,z} + e^{2\beta}r^{-2}A_{3,r}A_{3,z}]. \tag{54}$$

Here $A_3$ is the true magnetic component of the electromagnetic 4-potential and the superscript $m$ represents magnetovac case.

If one now compares the stationary gravitational equations (46) - (49) of the metric (45) with the magnetostatic field equations (51) - (54) corresponding to the metric (50), one obtains

$$\beta = u, \quad A_3 = i\omega, \quad \gamma^m = 4\gamma^s \tag{55}$$

It appears that magnetostatic solutions of EM equations can be constructed from the stationary vacuum solutions of Einstein's field equations and vice versa via the transformations (55). The imaginary quantity '$i$' in Eq. (55) can be eliminated by the parameter change technique of Refs. [20, 21].

It is evident from the transformation (55), that the magnetic potential $A_3$ can be obtained once $\omega$ is known. Here we encounter a problem for direct substitution of $\omega$ in equation (55) to evaluate $A_3$ since the metric (1) does not contain any term involving $\omega$. However, the problem of finding $\omega$ from the solutions of BZ can be solved following the prescription of Chaudhuri [3] discussed below:

It has been shown in Ref. [3] that for the stationary metric

$$ds^2 = f^{-1}\left[e^{2\gamma^s}\left(dr^2 + dz^2\right) + r^2 d\phi^2\right] - f(dt - \omega d\phi)^2 \quad , \tag{56}$$

The metric coefficients $f$ and $\omega$ are related to $g_{ab}$ of the metric (1) by

$$f = -g_{11} , \tag{57}$$

$$\omega = -(2g_{11})^{-1}(g_{12} + g_{21}) \tag{58}$$

In the case of a diagonal metric, $(\Omega)_{12} = (\Omega)_{21}$, and we have $g_{12} = g_{21}$. One thus obtains

$$\omega = -g_{12}(g_{11})^{-1} , \tag{59}$$

and $\qquad g_{12} = g_{21} = f\omega . \tag{60}$

The other metric coefficient $g_{22}$ is expressed as

$$g_{22} = K^2 f^{-1}(x^2 - 1)(1 - y^2) - f\omega^2 \tag{61}$$

Under the coordinate transformation (25), it is a simple task to establish from the above relations (57), (60) and (61) that $\det g_0 = -r^2$. This satisfies the supplementary condition (2).

Considering the line elements (45) & (56) and from equations (57) & (59) the expressions for $e^u$ and $\omega$ are obtained as

$$e^u = \frac{[2c_1^2\{x^2+(1-2y^2)\}-(c_3^2+c_4^2)(x^2-1)]}{[2c_1^2\{x^2-(1+2y^2)\}-c_3^2(x-1)^2-c_4^2(x+1)^2]} \tag{62}$$

$$\omega = 4K \frac{[c_1c_3\{x^2-x(1-y^2)-y^2\}-c_1c_4\{x^2+x(1-y^2)-y^2\}]}{[2c_1^2\{x^2+(1-2y^2)\}-(c_3^2+c_4^2)(x^2-1)]} \tag{63}$$

Using the transformation (55), physically realistic magnetostatic solutions are generated from equations (62) and (63) with the substitution $c_1 \to ic_1$. The results are,

$$e^{2\beta} = \left[\frac{[(c_3^2+c_4^2)(x^2-1)]+2c_1^2\{x^2+(1-2y^2)\}}{[c_3^2(x-1)^2+c_4^2(x+1)^2]+2c_1^2\{x^2-(1+2y^2)\}}\right]^2 \tag{64}$$

$$A_3 = 4K \frac{[c_1c_3\{x^2-x(1-y^2)-y^2\}-c_1c_4\{x^2+x(1-y^2)-y^2\}]}{[2c_1^2\{x^2+(1-2y^2)\}+(c_3^2+c_4^2)(x^2-1)]} \tag{65}$$

The asymptotic expansions of $e^{2\beta}$ and $A_3$ are obtained as

$$e^{2\beta} = 1 - 4\left(\frac{a_2}{a_1}\right)\frac{1}{x} + 8\frac{1}{a_1^2}\left(a_2^2+2a_3\right)\frac{1}{x^2} + \ldots \tag{66}$$

$$A_3 = 4Kc_1\left[\left(\frac{a_4}{a_1}\right)\frac{(1-y^2)}{x} - \left(\frac{a_5}{a_1^2}\right)\{a_6+(4c_1^2-a_1^2)y^2\}\frac{1}{x^2} + \ldots\right] \tag{67}$$

where,

$$a_1 = c_4^2 + c_3^2 + 2c_1^2, \quad a_2 = c_4^2 - c_3^2, \quad a_3 = c_1^4 - c_4^2 c_3^2$$

$$a_4 = c_4 + c_3, \quad a_5 = c_4 - c_3, \quad a_6 = c_4^2 + c_3^2 - 2c_1^2$$

(68)

The generated solutions are thus asymptotically flat.

Under the coordinate transformations defined by [16]

$$Kx = R - m, \quad \text{and} \quad y = \cos\theta, \quad (m \text{ is a constant}) \tag{69}$$

and with proper adjustment of the constants, the asymptotic expansions of $e^{2\beta}$ and $A_3$ are found to be

$$e^{2\beta} = 1 - \frac{4a_2 K}{a_1 R} + 4K \left[ \frac{2(a_2^2 + 2a_3)}{a_1^2} - m\left(\frac{a_2}{a_1}\right) \right] \cdot \frac{1}{R^2} + \ldots \ldots \ldots \tag{70}$$

$$A_3 = 4K^2 c_1 \left(\frac{a_4}{a_1}\right) \frac{\sin^2\theta}{R} - K \left[ 2\left(\frac{c_1^2}{a_1}\right)(1 + \sin^2\theta) - \left(\frac{ma_4}{a_5} - 1\right)\sin^2\theta \right] \frac{1}{R^2} + \ldots \ldots \ldots$$

(71)

The mass ($M$) and the magnetic dipole moment ($\mu$) of the source are evaluated from the asymptotic expansions of $e^{2\beta}$ and $A_3$ (i.e. from eqs. (70) and (71)) as

$$M = 2K \frac{\left(c_4^2 - c_3^2\right)}{\left(c_4^2 + c_3^2 + 2c_1^2\right)} \tag{72}$$

$$\mu = 4K^2 \frac{c_1(c_4 + c_3)}{\left(c_4^2 + c_3^2 + 2c_1^2\right)} \tag{73}$$

With further substitution of $c_1 = \alpha$, $c_3 = \alpha^2$ and $c_4 = 1$ (here $\alpha$ is any parameter), one obtains from equations (72) and (73) the mass ($M$) and magnetic dipole moment ($\mu$) as

$$M = 2K \frac{(1 - \alpha^2)}{(1 + \alpha^2)}, \tag{74}$$

$$\mu = 4K^2 \frac{\alpha}{(1 + \alpha^2)}. \tag{75}$$

The parameter $\alpha$ is thus identified as the magnetic parameter since the dipole moment $\mu$ = 0 for $\alpha$ = 0.

If the magnetic parameter $\alpha$ is set equal to zero and we take $K = m$, the mass of the source becomes $M = 2m$. This does not represent Schwarzschild solution but refers to a mass with monopole moment $2m$. Bonnor [6] also constructed a similar solution having the same mass monopole moment $M = 2m$.

For $\alpha = 1$, the mass of the source $M = 0$ and the magnetic dipole moment $\mu = 2K^2$. We thus obtain the dipole moment of a mass less source. This may be interpreted as the existence of finite current loop at large distances.

From equation (64) and with the substitutions $c_1 = \alpha$, $c_3 = \alpha^2$, $c_4 = 1$, the expression for $e^{2\beta}$ becomes

$$e^{2\beta} = \left[\frac{(x^2-1)(1+\alpha^2)^2 + 4\alpha^2(1-y^2)}{\{(x+1) + \alpha^2(x-1)\}^2 - 4\alpha^2 y^2}\right]^2$$

(76)

Redefining the constants as

$$\frac{(1+\alpha^2)}{(1-\alpha^2)} = \frac{K}{m} \ , \quad \frac{2\alpha}{(1-\alpha^2)} = \frac{b}{m} \ , \text{ such that } K^2 = m^2 + b^2 \ , \tag{77}$$

one obtains

$$e^{2\beta} = \left[\frac{K^2(x^2-1)+b^2(1-y^2)}{(Kx+m)^2 - b^2 y^2}\right]^2 \tag{78}$$

Under the coordinate transformation (69), a simple calculation gives

$$e^{2\beta} = \left[\frac{R^2 - 2mR - b^2 \cos^2\theta}{R^2 - b^2 \cos^2\theta}\right]^2$$

(79)

Equation (79) is the Bonnor's magnetic dipole solution [6]. Thus Bonnor's solution is exactly reproduced from our generated solutions using the techniques discussed above.

# 5 CONCLUSION:

A new technique is followed in the paper to construct static magnetic dipole solutions of Einstein-Maxwell field equations from the stationary gravitational soliton solutions of a Kerr body. The Kerr space-time is generated using the soliton technique of BZ. Although a number of static magnetic dipole solutions exist in the literature, [we mention here only a few of these such as 2, 6, 9, 19, 22 - 24], but the construction of magnetostatic solutions directly from stationary gravitational two-soliton solutions are not much discussed earlier. The transformations are straightforward and the results are obtained in a compact form. Particularly the use of the transformation presented in [3] makes the magnetostatic solution generation process from soliton solutions much simpler. In this respect the method followed in the paper is an interesting one and might be added to the list of new solution generating techniques of Einstein-Maxwell field equations.

Using the soliton technique of BZ, Kerr metric is derived in Eqs. (40) – (42). We make use of the transformations presented in Refs. [2, 3] and constructed the magnetostatic solutions in Eqs. (64) – (65). The generated solutions are found to be asymptotically flat containing monopole, dipole and other higher mass multipole moments. The mass and the magnetic dipole moment of the source are evaluated (equations (72) – (73)). For a particular value of the magnetic parameter, one obtains the magnetic dipole moment of a mass less source. This represents a finite current loop at large distances. Further it is worth mentioning that with proper adjustment of the constants and under the coordinate transformations (69), Bonnor's magnetic dipole solutions [6] are exactly reproduced from the results obtained by us (see Eq. (79)). It thus establishes the viability of the solution generating techniques presented in the paper.